\documentstyle[aps,prl,tighten,floats,psfig]{revtex}
\begin{document}
\draft
\binoppenalty10000 \relpenalty10000 \clubpenalty10000 \widowpenalty10000

\twocolumn[\hsize\textwidth\columnwidth\hsize
\csname@twocolumnfalse\endcsname

\title{Incommensurate spin fluctuations in the two-dimensional $t$-$t^\prime$-$U$ model}
\author{Adolfo Avella, Ferdinando Mancini and Dario Villani}
\address{Dipartimento di Scienze Fisiche {\em ''E.R. Caianiello''} - Unit\`a INFM di Salerno\\
Universit\`a degli Studi di Salerno, 84081 Baronissi (SA), Italy}
\date{July 24, 1997}
\maketitle

\begin{abstract} \widetext
Magnetic properties of the two-dimensional $t$-$t^{\prime }$-$U$ model are investigated by studying the static spin magnetic susceptibility as a function of momentum for various temperatures. The calculations are performed by means of the Composite Operator Method in the static approximation. By increasing the value of the $t^{\prime }$ parameter the magnetic scattering in the reciprocal space evolves to an isotropic structure. It is shown that the results are in qualitative agreement with the experimental situation observed in $LSCO$ and $YBCO$ compounds.
\end{abstract}
]

\narrowtext

The persistence of antiferromagnetic fluctuations in the
superconducting state of high-$T_c$ cuprates is one of the most striking
features in these materials. Indeed, it is widely accepted that in cuprate
materials there is a close relation between the unusual magnetic properties
and the occurrence of high temperature superconductivity, and that a
comprehension of the magnetic correlations in the normal state may be an
important step in the understanding the microscopic mechanism of pairing.

The knowledge of the wave-vector and energy dependencies of the spin
excitation spectrum is of the most importance in the attempt to build up an
appropriate theory for high-$T_c$ superconductivity\cite{Varma:1989}. The
dynamical spin susceptibility for cuprate materials has been investigated by
inelastic neutron scattering and {\em NMR} techniques. Neutron scattering
data on $La_{2-x}\left( Ba,Sr\right) _xCuO_4$ have shown\cite
{Thurston:1989,Shirane:1989,Birgenau:1988,Cheong:1991,Shirane:1994,Yamada:1995,Yamada:1996,Petit:1996}
that away from half-filling the magnetic Bragg peak in the dynamical
structure factor $S\left( {\bf k},\omega \right) $ broadens and develops a
structure with four peaks located at $[(1\pm \delta )\pi ,\pi ]$ and $[\pi
,(1\pm \delta )\pi ]$. Also, very recent neutron scattering experiments\cite
{Dai:1997} have solved the commensurate peak structure issue in $%
YBa_2Cu_3O_{6+y}$. P. Dai {\it et al.}\cite{Dai:1997} have shown that the
magnetic response in $YBa_2Cu_3O_{6.6}$ is complex with incommensurate
fluctuations at low temperatures for energies below the commensurate
resonance\cite{Dai:1996}. The low frequency spin fluctuations change from
commensurate to incommensurate on cooling with the incommensuration first
appearing at temperatures somewhat above $T_c$. The same behavior has been
observed in $Bi_2Sr_2CaCu_2O_{8-x}$\cite{Dai:1997}.

In a previous paper\cite{Mancini:1997}, in the context of a single-band
Hubbard model, we advanced a theoretical prediction; namely, it was claimed
a close relation between superconductivity and incommensurate magnetism in
some high-$T_c$ cuprates due to the reported proportionality between the
calculated amplitude of incommensurability and the experimental
superconducting critical temperature for $La_{2-x}Sr_xCuO_4$ over the whole
phase diagram. In the present experimental context, where the
incommensurability seems to be a common feature for all cuprate
superconductors, it results natural to revisit the analysis of the spin
fluctuations spectrum by adding a finite diagonal hopping term $t^{\prime }$
to the original Hubbard Hamiltonian. Infact, the addition of a $t^{\prime }$
bare parameter has often been suggested to handle the complexity of the
experimental situation for the cuprates\cite{Duffy:1995,Trapper:1996}.
Moreover, the next nearest neighbor hopping parameter $t^{\prime }$ emerges
from various reduction procedures as the single parameter, which carries, at
the level of the single-band description, the information about the crystal
structure outside the $CuO_2$ planes and thus differentiates between the
various cuprates\cite{Feiner:1996,Feiner:1996a}.

In this letter, we focus the attention on the momentum dependence of the
static spin magnetic susceptibility $\chi \left( {\bf k}\right) $, because
this quantity provides strict information about the spatial range of the
magnetic correlations. We show that the $t$-$t^{\prime }$-$U$ model presents
an incommensurate phase at finite doping that disappears by increasing
temperature. In addition, the next-nearest neighbor hopping term drives the
evolution of the magnetic scattering to an isotropic structure in the
reciprocal space.

The $t$-$t^{\prime }$-$U$ model is described by the Hamiltonian 
\begin{equation}
H=\sum_{ij}t_{ij}c^{\dagger }\left( i\right) c\left( j\right)
+U\sum_in_{\uparrow }\left( i\right) n_{\downarrow }\left( i\right) -\mu
\sum_in\left( i\right)
\end{equation}
where we use a standard notation. In the hopping matrix $t_{ij}$ we have
retained the terms up to the next-nearest neighbors situated along the
plaquette diagonals. In the static approximation\cite{Mancini:1995a}, the $%
{\bf k}$-dependent susceptibility $\chi ({\bf k})$ can be written as\cite
{Mancini:1997} 
\begin{equation}
\chi \left( {\bf k}\right) =\sum_{i,j=1}^2\chi _{ij}\left( {\bf k}\right)
\end{equation}
where 
\begin{equation}
\chi _{ij}\left( {\bf k}\right) =\int {\frac{d{\bf p}}{(2\pi )^2}}{\frac{%
f\left[ E_i({\bf k}+{\bf p})\right] -f\left[ E_j({\bf p})\right] }{E_i\left( 
{\bf k}+{\bf p}\right) -E_j\left( {\bf p}\right) }}K_{ij}\left( {\bf k},{\bf %
p}\right)
\end{equation}
$E_1\left( {\bf k}\right) $ and $E_2\left( {\bf k}\right) $ are the upper
and the lower Hubbard subbands, $K_{ij}\left( {\bf k},{\bf p}\right) $ are
expressed in terms of the spectral intensities. The explicit expressions for
these quantities have been computed in the framework of the {\em COM}\cite
{Mancini:1995a,Avella:1997,Mancini:1995b}. The term $\chi _{inter}=\chi
_{12}+\chi _{21}$ describes the propagation of a spin accompanied by a spin
excitation between the two bands $E_1\left( {\bf k}\right) $ and $E_2\left( 
{\bf k}\right) $; while the two terms $\chi _{11}$ and $\chi _{22}$ describe
the propagation with a subsequent intraband spin excitation.

%TCIMACRO{
%\TeXButton{Figure: fig1}{\begin{figure}[tb]
%\centerline{\psfig{figure=fig1.ps,angle=270,width=8cm}}
%\caption{The static susceptibility $\chi\left({\bf k}\right)$ for various temperatures. $U=4$, $t^{\prime }=0$ and $n=0.97$. In panels $\left( a\right) $ and $\left( b\right) $ are reported the lines ${\bf k}=\left(\pi,k\right)$ and ${\bf k}=\left( k,k\right)$, respectively.}
%\label{fig1}
%\end{figure}}}
%BeginExpansion
\begin{figure}[tb]
%\centerline{\psfig{figure=fig1.ps,angle=270,width=8cm}}
\caption{The static susceptibility $\chi\left({\bf k}\right)$ for various temperatures. $U=4$, $t^{\prime }=0$ and $n=0.97$. In panels $\left( a\right) $ and $\left( b\right) $ are reported the lines ${\bf k}=\left(\pi,k\right)$ and ${\bf k}=\left( k,k\right)$, respectively.}
\label{fig1}
\end{figure}
%EndExpansion

%TCIMACRO{
%\TeXButton{Figure: fig2}{\begin{figure}[tb]
%\centerline{\psfig{figure=fig2.ps,angle=270,width=8cm}}
%\caption{Surface plot in reciprocal space of $\chi\left({\bf k}\right)$ for $U=4$, $t^{\prime }=0$, $n=0.97$ and $T=0.01$}
%\label{fig2}
%\end{figure}}}
%BeginExpansion
\begin{figure}[tb]
%\centerline{\psfig{figure=fig2.ps,angle=270,width=8cm}}
\caption{Surface plot in reciprocal space of $\chi\left({\bf k}\right)$ for $U=4$, $t^{\prime }=0$, $n=0.97$ and $T=0.01$}
\label{fig2}
\end{figure}
%EndExpansion

%TCIMACRO{
%\TeXButton{Figure: fig3}{\begin{figure}[tb]
%\centerline{\psfig{figure=fig3.ps,angle=270,width=8cm}}
%\caption{Contour plot in reciprocal space of $\chi\left({\bf k}\right)$ for $U=4$, $t^{\prime }=0$, $n=0.97$ and $T=0.01$}
%\label{fig3}
%\end{figure}}}
%BeginExpansion
\begin{figure}[tb]
%\centerline{\psfig{figure=fig3.ps,angle=270,width=8cm}}
\caption{Contour plot in reciprocal space of $\chi\left({\bf k}\right)$ for $U=4$, $t^{\prime }=0$, $n=0.97$ and $T=0.01$}
\label{fig3}
\end{figure}
%EndExpansion

In Fig.~\ref{fig1} we present the static susceptibility $\chi \left( {\bf k}%
\right) $ for various temperatures. The choice for the parameters is $U=4$, $%
t^{\prime }=0$ and the particle filling has been fixed as $n=0.97$. In this
letter all the energies are expressed in unity of $t$. In panels $\left(
a\right) $ and $\left( b\right) $ $\chi \left( {\bf k}\right) $ is reported
along the line ${\bf k=}\left( \pi ,k\right) $ and the line ${\bf k=}\left(
k,k\right) $, respectively. In both cases by increasing the temperature the
incommensurate double-peak structure becomes a broad maximum centered at $%
\left( \pi ,\pi \right) $. Along the $\left( k,k\right) $ line the intensity
at incommensurate positions is lower than the one along $\left( \pi
,k\right) $. This can be clearly observed in Figs.~\ref{fig2} and~\ref{fig3}
where four well-resolved incommensurate peaks are located at $\left[ \left(
1\pm \delta \right) \pi ,\pi \right] $ and $\left[ \pi ,\left( 1\pm \delta
\right) \pi \right] $. The important features of the data are:

\begin{enumerate}
\item  the overall square symmetry of the scattering with the sides of the
square parallel to the $\left( k,k\right) $ and to the $\left( k,-k\right) $
lines;

\item  the accumulation of intensity near the corners of the square.
\end{enumerate}

These features reproduce the experimental situation for $La_{2-x}\left(
Ba,Sr\right) _xCuO_4$\cite
{Thurston:1989,Shirane:1989,Birgenau:1988,Cheong:1991,Shirane:1994,Yamada:1995,Yamada:1996,Petit:1996}%
.

%TCIMACRO{
%\TeXButton{Figure: fig4}{\begin{figure}[tb]
%\centerline{\psfig{figure=fig4.ps,angle=270,width=8cm}}
%\caption{Surface plot in reciprocal space of $\chi\left({\bf k}\right)$ for $U=4$, $t^{\prime }=-0.1$, $n=0.97$ and $T=0.01$}
%\label{fig4}
%\end{figure}}}
%BeginExpansion
\begin{figure}[tb]
%\centerline{\psfig{figure=fig4.ps,angle=270,width=8cm}}
\caption{Surface plot in reciprocal space of $\chi\left({\bf k}\right)$ for $U=4$, $t^{\prime }=-0.1$, $n=0.97$ and $T=0.01$}
\label{fig4}
\end{figure}
%EndExpansion

%TCIMACRO{
%\TeXButton{Figure: fig5}{\begin{figure}[tb]
%\centerline{\psfig{figure=fig5.ps,angle=270,width=8cm}}
%\caption{Contour plot in reciprocal space of $\chi\left({\bf k}\right)$ for $U=4$, $t^{\prime }=-0.1$, $n=0.97$ and $T=0.01$}
%\label{fig5}
%\end{figure}}}
%BeginExpansion
\begin{figure}[tb]
%\centerline{\psfig{figure=fig5.ps,angle=270,width=8cm}}
\caption{Contour plot in reciprocal space of $\chi\left({\bf k}\right)$ for $U=4$, $t^{\prime }=-0.1$, $n=0.97$ and $T=0.01$}
\label{fig5}
\end{figure}
%EndExpansion

In Figs.~\ref{fig4} and~\ref{fig5} we report the results for $t^{\prime
}=-0.1$. In this case incommensurate magnetic resonance modes are
isotropically distributed around the $\left( \pi ,\pi \right) $ point. The
occurrence of such an evolution of the magnetic fluctuations can be related
to the spreading of the nesting vector in the momentum space. That is, the
evolution from a pseudo-nested to a roughly circular hole-like Fermi
surface. Infact, the shape and, in particular, the bending of the Fermi
surface are strongly dependent on the value of the $t^{\prime }$ parameter,
which leads for fixed values of filling $n$ and of the $U$ parameter to a
real rotation of the Fermi surface\cite{Avella:1997}.

%TCIMACRO{
%\TeXButton{Figure: fig6}{\begin{figure}[tb]
%\centerline{\psfig{figure=fig6.ps,angle=270,width=8cm}}
%\caption{The static susceptibility $\chi\left({\bf k}\right)$ for various temperatures. $U=4$, $t^{\prime }=-0.1$ and $n=0.97$. In panels $\left( a\right) $ and $\left( b\right) $ are reported the lines ${\bf k}=\left(\pi,k\right)$ and ${\bf k}=\left( k,k\right)$, respectively.}
%\label{fig6}
%\end{figure}}}
%BeginExpansion
\begin{figure}[tb]
%\centerline{\psfig{figure=fig6.ps,angle=270,width=8cm}}
\caption{The static susceptibility $\chi\left({\bf k}\right)$ for various temperatures. $U=4$, $t^{\prime }=-0.1$ and $n=0.97$. In panels $\left( a\right) $ and $\left( b\right) $ are reported the lines ${\bf k}=\left(\pi,k\right)$ and ${\bf k}=\left( k,k\right)$, respectively.}
\label{fig6}
\end{figure}
%EndExpansion

Besides, the intensity of the incommensurate peaks along the $\left(
k,k\right) $ line is now comparable to that along $\left( \pi ,k\right) $
(Fig.~\ref{fig6}). In this case the structure along the diagonals is not
related to a saddle point topology as in $LSCO$, but to a really increased
scattering as it has been observed in $YBa_2Cu_3O_{6.6}$\cite{Dai:1997}.
Again, by increasing temperature the commensurability is recovered. An
isotropic magnetic scattering, as we have found for $t^{\prime }$ different
from zero, should be detected for $YBCO$ once extended scans over the whole
Brillouin zone will be made. Indeed, this prediction is supported by {\em %
ARPES} data for $YBCO$\cite{Shen:1995}, which seem to exclude an underlying
symmetry for the in-plane magnetic scattering identical to that of $LSCO$,
but $\pi /4$ rotated\cite{Dai:1997}.

Performed theoretical calculations show that by increasing further $%
t^{\prime }$ the magnetic scattering processes are suppressed at all wave
vectors so that we have a featureless susceptibility with no clearly
detectable peaks over all the Brillouin zone.

In conclusion, we have shown that the $t$-$t^{\prime }$-$U$ model presents
an incommensurate phase at finite doping and that the magnetic scattering
become isotropic as $t^{\prime }$ is increased. The occurrence of such an
evolution of the magnetic fluctuations can be related to the spreading of
the nesting vector in the momentum space. That is, the evolution in the
dynamical response of a spatially uniform electron liquid from a
pseudo-nested to a roughly circular hole-like Fermi surface.

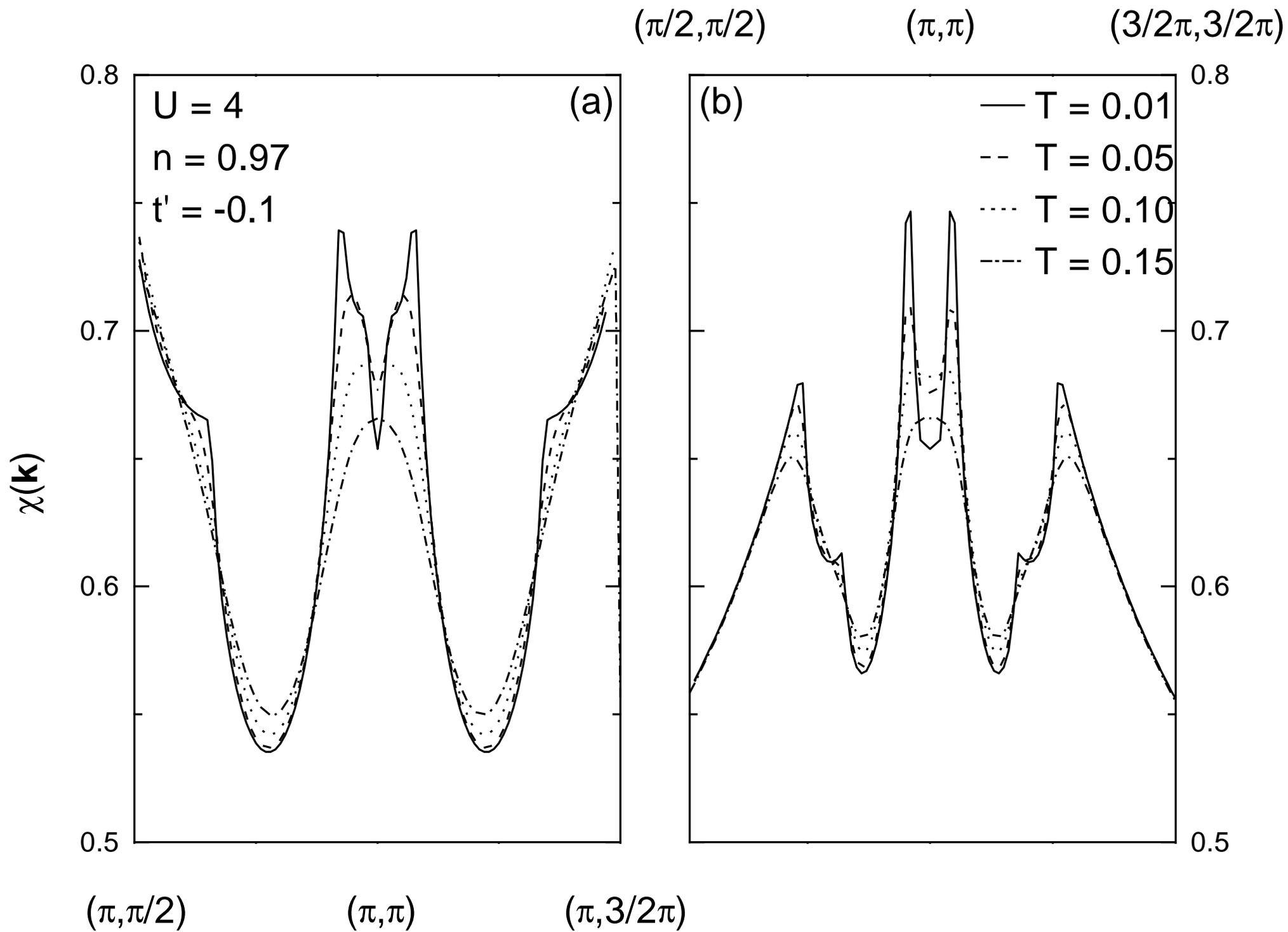
\end{document}